\documentstyle[12pt]{article}
\newcommand{\slp}{\raise.15ex\hbox{$/$}\kern-.57em\hbox{$\partial$}}
\newcommand{\sla}{\raise.15ex\hbox{$/$}\kern-.57em\hbox{$a$}}
\newcommand{\slA}{\raise.15ex\hbox{$/$}\kern-.57em\hbox{$A$}}
\newcommand{\slB}{\raise.15ex\hbox{$/$}\kern-.57em\hbox{$B$}}
\newcommand{\slb}{\raise.15ex\hbox{$/$}\kern-.57em\hbox{$b$}}
\newcommand{\slW}{\raise.15ex\hbox{$/$}\kern-.57em\hbox{$W$}}
\newcommand{\be}{\begin{equation}}
\newcommand{\ee}
{\end{equation}}
\newcommand{\bear}{\begin{eqnarray}}
\newcommand{\ear}{\end{eqnarray}}

\newcommand{\D}{\cal D}

\begin{document}
\begin{flushright}
HD--THEP--96--28\\
La Plata-Th 96/08
\end{flushright}
\quad\\
\vspace{1.8cm}
\begin{center}
{\bf\LARGE
On the physical Hilbert space of QCD$_2$}\\
\medskip
{\bf\LARGE in the decoupled
formulation}\\
\vspace{1cm}
D.C. Cabra\\
Departamento de F\'\i sica, Universidad Nacional de la Plata\\
C.C. 67, (1900) La Plata, Argentina\\
\bigskip
K.D. Rothe\\
Institut  f\"ur Theoretische Physik\\
Universit\"at Heidelberg\\
Philosophenweg 16, D-69120
Heidelberg\\
\end{center}
\vspace{2.0cm}
\begin{abstract}
We consider the QCD$_2$ partition function in the non-local, decoupled
formulation and systematically establish which subset of the nilpotent
Noether charges is required to vanish on the physical states. The
implications for the Hilbert space structure are also examined.
\end{abstract}
\newpage
\section{Introduction}
The formulation of two-dimensional quantum chromodynamics (QCD$_2$) of
massless
fermions in terms of decoupled fermions, ghosts and positive and negative
 level Wess-Zumino fields \cite{FNS,AA1,AA2} has provided interesting insight
into some non-perturbative properties of this theory. Two representations
of the corresponding decoupled  partition function, referred to as
``local'' and ``non-local''  representations \cite{AA1,AA2}, have been
considered. In the ``local'' formulation, the original restriction
 of the ``observables'' to the gauge invariant
subspace of the Hilbert space
is replaced in the light-cone gauge $A_+\equiv A_0+A_1=0$ by the
requirement that ``observables'' commute with two BRST charges
\cite{CRS,RST}.

When passing to the ``non-local'' formulation, one
expects to pick up one
additional BRST condition associated with
the change of variable involved
in the transition. De facto one finds, however, more than three nilpotent
 charges, which are moreover non-commuting.  This raises the question
as to which of these charges are  required to annihilate the physical
states. This question has recently been addressed in the context
of quantum mechanical toy models and the \underbar{local}
decoupled formulation
of QCD$_2$, in ref.\cite{RST}, where criteria have also been given for
establishing which BRST conditions should actually be imposed.

The primary aim of the present paper is to examine this question for
the case of QCD$_2$ in the non-local decoupled formulation. As we show
in section 2, not all of the nilpotent charges obtained in ref.\cite{CRS}
are required to vanish on the physical Hilbert space ${\cal H}_{\rm phys}$.
In section 3 we solve the corresponding  cohomology problem in the ghost
number zero sector by showing that the BRST conditions which are actually
to be imposed  are implemented by the gauge invariant observables
of the theory.

In section 4, we then discuss the conformal ``sector'' of the
factorized non-local partition function. In ref.\cite{AR1}
the QCD$_2$ ground state was taken to lie in this sector.  It was thereby
concluded that in  the case of one flavor  and gauge group  $SU(2)$ the
QCD$_2$ ground state is two-fold degenerate. We generalize
this statement to the case of $SU(N)$-color.

Section 5 summarizes our results.

\section{BRST constraints}

We reconsider here the BRST analysis of ref.\cite{CRS}. We discuss
separately the ``local'' \cite{FNS,AA1} and ``non-local'' formulations
\cite{AA1,AA2}.

\subsection{Local formulation}

The QCD$_2$ partition function is given by
\be\label{2.1}
Z=\int{{\D}} A_\mu\int{{\D}}\psi{{\D}}\bar\psi
\exp[-i\int\frac{1}{4}tr  F_{\mu\nu}F^{\mu\nu}]
\exp [i\int\bar\psi(i{\raise.15ex\hbox{$/$}\kern-.57em\hbox{$\partial$}}
+e{\raise.15ex\hbox{$/$}\kern-.57em\hbox{$A$}})\psi], \ee
where $F_{\mu\nu}$ is the chromoelectric field strength tensor.
Going to the light-cone gauge $A_+=0$, parametrizing $A_-$ as
\be\label{2.2}
eA_-=Vi\partial_- V^{-1},\ee
and performing a chiral rotation, $\psi_2=V\psi^{(0)}_2$,
 one arrives at the decoupled
partition function (for details see for instance ref.\cite{CRS})
\be\label{2.3}
Z=Z^{(0)}_F Z^{(0)}_{gh} Z_V, \ee
where $Z^{(0)}_F,\ Z^{(0)}_{gh}$ are the partition functions of
massless free fermions and ghosts, respectively,
\bear\label{2.4}
Z^{(0)}_F&=&\int {\D}\psi{\D}\bar\psi\exp[i\int\bar\psi^{(0)}i\slp
\psi^{(0)}],\nonumber\\
Z^{(0)}_{gh}&=&\int{\D}({\rm ghosts})\exp\left \{i\int[ b^{(0)}_+i\partial_-
c^{(0)}_+ + b_-^{(0)}i\partial_+ c_-^{(0)}]\right\},\ear
and
\be\label{2.5}
Z_V=\int{\D}V\exp\left\{-i(1+c_V)\Gamma[V]+\frac{i}{8e^2}
\int d^2 xtr [\partial_+(Vi\partial_- V^{-1})]^2\right\}.\ee
Here $\Gamma[V]$ is the usual Wess-Zumino-Witten (WZW) functional
\cite{WZW}, and $c_V$ is the Casimir of the gauge group.

As shown in ref.\cite{CRS} this decoupled partition function
exhibits two BRST symmetries implying the existence of two nilpotent
Noether charges. The corresponding BRST currents are \cite{CRS}
\be\label{2.6}
{\cal J}_{\pm}=tr  c_\pm^{(0)}\left[\Omega_\pm-\frac{1}{2}
\left\{b_\pm^{(0)},c_\pm^{(0)}\right\}\right], \ee
where
\bear\label{2.7}
\Omega_-&=&-\frac{1}{4e^2}{\D}_-(V)\partial_+(Vi\partial_- V^{-1})-
(1+c_V)J_-(V)+j_- , \nonumber\\
\Omega_+&=&-\frac{1}{4e^2}{\D}_+(V)\partial_-(V^{-1}i\partial_+ V)
-(1+c_V) J_+(V)+j_+.\ear
Here ${\D}_\pm(V)$ are the covariant derivatives
\bear\label{2.8}
{\D}_+(V)&=&\partial_++[V^{-1}\partial_+V,\ ] , \nonumber\\
{\D}_-(V)&=&\partial_-+[V\partial_- V^{-1},\ ]\ear
and use has been made of the identity
\be\label{2.9}
V^{-1}\left[\partial^2_+(V\partial_- V^{-1})\right]V={\D}
_+(V)\partial_-(V^{-1}\partial_+ V), \ee
in order to write the BRST currents of ref.\cite{CRS} in symmetrical form.
$J_\pm(V)$ and $j_\pm$ are the currents
\bear\label{2.10}
&&J_+(V)=\frac{1}{4\pi} V^{-1} i\partial_+ V,\nonumber\\
&&J_-(V)=\frac{1}{4\pi}
 Vi\partial_- V^{-1}, \nonumber\\
&&j_-=\psi^{(0)}_1 \psi^{(0)\dagger}_1+\left\{ b_-^{(0)}
, c_-^{(0)}\right\},\nonumber\\
&&j_+=\psi_2^{(0)}\psi_2^{(0)\dagger}+\left\{ b^{(0)}_+,c^{(0)}_+\right\}
.\ear

The gauge invariance of the observables in the original formulation
(\ref{2.1})  is replaced in the decoupled picture by the requirement that
the physical operators commute with the BRST charges $Q_\pm$ associated
with ${\cal J}_\pm$ (see \cite{RST} for proof). In the ghost number zero
sector
this implies that $\Omega_\pm$  defined by (\ref{2.7})
are constrained to vanish on the physical subspace:
\be\label{2.11}
\Omega_\pm\approx 0.\ee
As shown in ref.\cite{KS}, this property can independently be established
by appropriately gauging the action in the  decoupled partition function.

\subsection{Non-local formulation}

The main objective of this paper is to trace the fate of the BRST
conditions
 of the local formulation, when going over to the so-called \cite{AA1}
``non-local'' formulation, and to establish from first principles
further  BRST conditions that may have to be imposed in order to ensure
equivalence of this  formulation with the local one.

In order to make the discussion self-contained, we repeat here the
essential steps leading to the non-local
formulation \cite{AA1,AA2,CRS}. We rewrite the partition function $Z_V$ given
in (\ref{2.5}) by making use of the identity

\bear\label{2.11a}
\lefteqn{\exp\left\{\frac{i}{4e^2}\int tr \frac{1}{2}
[\partial_+(V i\partial_- V^{-1})]^2\right\}}
\nonumber\\
& & =\int{\D}
E\exp\left\{-i\int tr \left[\frac{1}{2} E^2+\frac{E}{2e}
\partial_+(V i\partial_- V^{-1})\right]\right\}.\ear
Making the change of variable

\be\label{2.12}
\partial_+ E=\lambda\beta^{-1} i\partial_+\beta\ ,\ \lambda
=\left(\frac{1+ c_V}{2\pi}\right)e , \ee
we have for the corresponding change in the integration measure
\be\label{2.13}
{\D} E=e^{-ic_V\Gamma[\beta]}{\D}\beta.\ee
Making use of the Polyakov-Wiegmann identity \cite{PW}
\be\label{2.14}
\Gamma[gh]=\Gamma[g]+\Gamma[h]+\frac{1}{4\pi}
\int d^2xtr \left(g^{-1} \partial_+ g h\partial_-h^{-1}\right),\ee
and defining the new variable
\be\label{2.15}
\tilde V=\beta V, \ee
one then arrives at a decoupled non-local form of the partition
function \cite{AA1,CRS}:
\be\label{2.16}
Z=Z^{(0)}_F Z^{(0)}_{gh} Z_{\tilde V} Z_\beta ,\ee
where
\be\label{2.17}
Z_{\tilde V}=\int{\D} \tilde V e^{-i(1+c_V)\Gamma[\tilde V]},\ee
and
\be\label{2.18}
Z_\beta=\int{\D}\beta e^{i\Gamma[\beta]+i\lambda^2\int\frac{1}{2} tr
[\partial^{-1}_+(\beta^{-1}\partial_+\beta)]^2}.\ee
We now investigate the BRST conditions to be imposed on the
physical states in this formulation.\\
\bigskip

{\it a) BRST condition associated with the change of variable
 $E\to\beta$}\\

We begin by showing that the change of variable (\ref{2.12}) leads to
a BRST condition on the physical states. We follow the procedure
outlined in ref.\cite{Ba}.

In order to implement the change of variable (\ref{2.12}),
we introduce in (\ref{2.11a}) the identity
\be\label{2.19}
\lambda\int{\D}\beta det{\D}_+(\beta)\delta\left[ \partial_+E
-\lambda \beta^{-1}i\partial_+\beta\right] =1 , \ee
in order to rewrite (\ref{2.5}) in the form
\be\label{2.20}
Z_V=\int{\D} V\int{\D}E {\D}\beta\int{\D}\rho\int{\D}(ghosts)
e^{iS[E,V]} e^{i\Delta S[E,\beta,\rho,ghosts]},\ee
where
\be\label{2.21}
S[E,V]=-(1+c_V)\Gamma[V]-\int d^2x[\frac{1}{2} E^2+\frac{E}{2e}
\partial_+(Vi\partial_-V^{-1})]\ee
and
\be\label{2.22}
\Delta S[E,\beta,\rho,ghosts]=\int  d^2x\{\rho(\partial_+E-
\lambda\beta^{-1}i\partial_+\beta)+\hat b_-i{\D}_+(\beta)\hat c_-
\} . \ee
Here we have made use of the Fourier representation of the
$\delta$ functional in (\ref{2.19}), and the representation
of the adjoint determinant $det{\D}_+(\beta)$ in terms of ghosts:
\be\label{2.23}
det{\D}_+(\beta)=\int{\D}\hat b{\D}\hat c e^{i\int \hat b i{\D}_+(\beta)\hat c}.
\ee
The effective action is seen to be invariant under the
transformation

\bear\label{2.24}
~&~&\delta V=0,\quad \delta E=0,\quad \delta\rho=0, \nonumber\\
~&~&\delta\hat b_- =\lambda\rho,\quad \delta\hat c_-=-\frac{1}
{2}\{\hat c_-,\hat c_-\}, \nonumber\\
~&~&\beta^{-1}\delta\beta=\hat c_-.\ear
One readily checks that these transformations
are off-shell nilpotent. We further observe that
$\Delta S$ in (\ref{2.22}) can be written as
\be\label{2.25}
\Delta S=\frac{1}{\lambda}\delta[\hat b_-(\partial_+E-\lambda\beta^{-1}i\partial
_+\beta)].\ee
Hence $\Delta S$ is BRST exact. From here we infer that the
actions $S[E,V]$ and $S[E,V]+\Delta S$ are equivalent
on the functionals which are invariant under the nilpotent
transformations (\ref{2.24}). Hence physical states must
be invariant under the transformations (\ref{2.24}).

In order to obtain the transformation laws (\ref{2.24}) in terms
of the variables of the non-local formulation, we make use of the
equations of motion for $\rho$ and $E$. The transformations
(\ref{2.24}) then reduce to
\bear\label{2.26}
&&\delta V=0,\nonumber\\
&&\delta\hat b_-=-\lambda^2\partial_+^{-2}(\beta^{-1}i\partial_+\beta)-
\frac{\lambda}{2e}(Vi\partial_-V^{-1}), \nonumber\\
&&\delta\hat c_-=-\frac{1}{2}\{\hat c_-,\hat c_-\}, \nonumber\\
&&\beta^{-1}\delta\beta=\hat c_-.\ear
We next decouple the ghosts by performing the change of variable
\bear\label{2.27}
&&\hat b_-\to\beta\hat b_-\beta^{-1}=:\hat b_-^{(0)},\nonumber\\
&&\hat c_-\to \beta\hat c_-\beta^{-1}=:\hat c^{(0)}. \ear

In terms of $\tilde V=\beta V$ and the
decoupled ghosts, the transformation laws then read
\bear\label{2.28}
\delta\tilde V\tilde V^{-1}&=&\hat c_-^{(0)}, \nonumber\\
\delta \beta\beta^{-1}&=&\hat c_-^{(0)}, \nonumber\\
\delta\hat c_-^{(0)}&=&\frac{1}{2}\{\hat c_-^{(0)},\hat c_-^{(0)}
\}, \nonumber\\
\delta \hat b^{(0)}_-&=&-\lambda^2\beta\partial_+^{-2}(\beta^{-1}
i\partial_+\beta)\beta^{-1}\nonumber\\
&&-\left(\frac{1+c_V}{4\pi}\right)\left[(\tilde V i\partial_-
\tilde V^{-1})-\beta i\partial_-\beta^{-1}\right]\nonumber\\
&&+ \{b^{(0)}_-,c^{(0)}_-\} + (\delta \hat b_-^{(0)})_{anom}, \ear
where the anomalous term
\be\label{2.29}
(\delta \hat b_-^{(0)})_{anom}=-\frac{c_V}{4\pi}\beta i\partial_-
\beta^{-1}, \ee
needs to be added to the semiclassical result in order to make
the transformation an invariance of the quantum action.
As has been shown in ref.\cite{CRS}, the transformation
laws (\ref{2.28}) lead to the (right-moving) BRST current
\be\label{2.30}
{\hat J}_-=tr {\hat c}_-^{(0)}\left({\hat\Omega}_--\frac{1}{2}
\{{\hat b}_-^{(0)}
,{\hat c}_-^{(0)}\}\right), \ee
with
\bear\label{2.31}
{\hat\Omega}_-&=&-\lambda^2\beta(\partial_+^{-2}(\beta^{-1}
i\partial_+\beta))\beta^{-1}+J_-(\beta)\nonumber\\
&&-(1+c_V)J_-(\tilde V)+\left\{{\hat b}_-^{(0)},\hat c_-^{(0)}
\right\}.\ear
Our deductive procedure shows that the corresponding
nilpotent charge $\hat Q_-$ must annihilate the physical
states:
\be\label{2.32}
\hat Q_-=0\quad {\rm on}\quad {\cal H}_{phys}.\ee
\bigskip

{\it b) Fate of the BRST condition $Q_+\approx 0$}\\

Making use of the identity (\ref{2.9}), we may rewrite
$\Omega_+$ in (\ref{2.7}) as
\be\label{2.33}
\Omega_+=-\frac{1}{4e^2}V^{-1}\left[\partial_+^2(Vi\partial_-
V^{-1})\right]V-(1+c_V)J_+(V)+j_+.\ee
Using the equation of motion for $E$ following from
(\ref{2.11a})
\be\label{2.34}
E=-\frac{1}{2e}\partial_+(Vi\partial_-V^{-1}), \ee
$\Omega_+$ takes the form
\be\label{2.35}
\Omega_+=\frac{1}{2}V^{-1}(\partial_+E)V-(1+c_V)J_+(V)+j_+.\ee
Making the change of variable (\ref{2.12}) and (\ref{2.15}),
we then obtain\footnote{For the sake of clarity we continue
to use the same notation for the constraints when expressed
in terms of the new variables.}
\be\label{2.36}
\Omega_+=-(1+c_V)J_+(\tilde V)+j_+, \ee
where $\tilde V=\beta V$. Comparing with eq.(\ref{3.13}) of ref.\cite{CRS},
we see that this is just $\tilde\Omega_+$ of ref.\cite{CRS}.
We conclude that the corresponding nilpotent charge
\be\label{2.37}
Q_+=\int dx^1tr c_+^{(0)}\left[-(1+c_V)J_+(\tilde V)
+j_+-\frac{1}{2}\left\{b_+^{(0)},c_+^{(0)}\right\}\right],
\ee
must annihilate the physical states, as was also required
in ref.\cite{CRS}.\\

\bigskip
{\it c) Fate of the BRST condition $Q_-\approx 0$}\\

In the case of the BRST charge $Q_+$, the symmetry transformations
in the $V$-fermion-ghost space giving rise to this conserved
charge could be trivially extended to the $E-V$-fermion-ghost
space. This is no longer true in the case of $Q_-$, where
the BRST symmetry for $E$ off-shell is maintained only at
the expense of the addition of a (commutator)
term (which vanishes for
$E$ ``on shell''). One is thereby led to a fairly
complicated expression for
$Q_-$ when expressed in terms of the variables $\beta, \tilde
V$ of the non-local formulation.

A more transparent result is obtained by performing the
similarity transformation

\be\label{2.38}
E'=-V^{-1}EV
\ee
and making the change of variables
\be\label{2.39}
\partial_-E'=\lambda\beta'i\partial_-{\beta'}^{-1}. \ee
Going through the same steps as outlined before,
one arrives at an alternative representation of the partition
function (\ref{2.16}),
\be\label{2.40}
Z=Z_F^{(0)}Z_{gh}^{(0)}Z_{\tilde V'}Z_{\beta'},\ee
where
\be\label{2.41}
Z_{{\tilde V}'}=\int{\D}\tilde V'e^{-i(1+C_V)\Gamma[\tilde V']},\ee
\be\label{2.42}
Z_{\beta'}=\int{\D}\beta'e^{i\Gamma[\beta']+
i\lambda^2\int \frac{1}{2}tr[\partial_-^{-1}(\beta'\partial_-{\beta'}^{-1})]^2}
\ee
and
\be\label{2.43}
\tilde V'=V\beta'. \ee
Note that $\beta'$ satisfies a different dynamics than
$\beta$ introduced previously.

It is convenient to rewrite $\Omega_-$ in eq. (\ref{2.7})
in the form
\be\label{2.44}
\Omega_- = -\frac{1}{4e^2}V\partial_-(V^{-1}[\partial_+(Vi\partial_-
V^{-1})]V)V^{-1}-(1+c_V)J_-(V)+j_-. 
\ee
Rewriting $\Omega_-$ in terms of $E'$ by making use of the equation
of motion
\be\label{2.45}
E'=-\frac{1}{2e}\partial_-(V^{-1}i\partial_+V)\ee
and making use of (\ref{2.39}), one arrives at
\be\label{2.46}
\Omega_-=-(1+c_V)J_-(\tilde V')+j_-.\ee
We conclude that the corresponding BRST charge
\be\label{2.47}
Q_-=\int dx^1tr c_-^{(0)}[-(1+c_V)J_-(\tilde V')
+j_--\frac{1}{2}\{b_-^{(0)},c_-^{(0)}\}]\ee
must annihilate the physical states.

Notice that expression (\ref{2.46}) formally resembles
$\tilde\Omega_-$ of ref.\cite{CRS}. Although $\tilde V$
in (\ref{2.37}) and $\tilde V'$ in (\ref{2.47}) obey
the same dynamics, they are, however, vinculated by different
constraints to the ``massive'' sector described in terms
of the group-valued fields $\beta$ and $\beta'$, respectively,
which in turn obey a different dynamics.
The constraint $\hat\Omega_+\approx 0$ associated with the
change of variable $E' \to \beta'$ is again obtained following the
previous systematic procedure, and one finds
\bear\label{2.48}
\hat\Omega_+&=&-\lambda^2{\beta'}^{-1}(\partial_-^{-2}
(\beta' i\partial_-{\beta'}^{-1}))\beta'+J_+(\beta')\nonumber\\
&&-(1+c_V)J_+(\tilde V')+\{\hat b_+^{(0)},\hat c_+^{(0)}\}.\ear
For similar reasons as before, the corresponding Noether charge
\be\label{2.49}
{\hat Q}_+=\int dx'tr{\hat c}_+^{(0)}\left[{\hat\Omega}_+-\frac{1}{2}
\left\{{\hat b}_+^{(0)},{\hat c}_+^{(0)}\right\}\right]\ee
must annihilate the physical states.

On the ghost number zero sector, the BRST conditions to be imposed
on the physical states are equivalent to requiring
\be\label{2.50}
\Omega_\pm\approx0,\quad\hat\Omega_\pm\approx0, \ee
with $\hat\Omega_\pm$ and $\Omega_\pm$ given by
eqs.(\ref{2.31}), (\ref{2.36}), (\ref{2.46}) and
(\ref{2.48}).

\section{The physical Hilbert space}
\setcounter{equation}{0}
In order to address the cohomology problem defining the
physical Hilbert space, we must express the constraints
in terms of canonically conjugate variables. To this end
we first rewrite the partition function $Z_\beta$ in (\ref{2.18})
in terms of an auxiliary field $B$ as follows:
\be\label{3.1}
Z_\beta=\int {\D}B{\D}\beta e^{iS[\beta,B]}, \ee
where
\be\label{3.2}
S[\beta,B]=\Gamma[\beta]+\int tr\left[\frac{1}{2}
(\partial_+B)^2+\lambda B\beta^{-1}i\partial_+\beta\right].\ee
Correspondingly we have for $Z_{\beta'}$ in (\ref{2.42})
\be\label{3.3}
Z_{\beta'}=\int {\D}B'{\D}\beta' e^{iS'[\beta',B']}, \ee
with
\be\label{3.4}
S'[\beta',B']=\Gamma[\beta']+\int tr\left[\frac{1}{2}
(\partial_-B')^2+\lambda B'\beta'i\partial_-{\beta'}^{-1}\right].\ee
We may then rewrite the constraints ${\hat\Omega}_\pm\approx 0$
in (\ref{2.31}) and (\ref{2.48}) as
\be\label{3.5}
\hat\Omega_-=\lambda\beta B\beta^{-1}+\frac{1}{4\pi}\beta i
\partial_-\beta^{-1}-\frac{(1+c_V)}{4\pi}\tilde V i\partial_-
\tilde V^{-1}+\{\hat b_-^{(0)},\hat c_-^{(0)}\},\ee
\be\label{3.6}
\hat\Omega_+=\lambda{\beta'}^{-1} B'\beta'+\frac{1}{4\pi}{\beta'}^{-1}
i\partial_+\beta'-\frac{(1+c_V)}{4\pi}\tilde {V'}^{-1} i\partial_+
\tilde V'+\{\hat b_+^{(0)},\hat c_+^{(0)}\}.\ee
Define (tilde stands for ``transpose'')
\bear\label{3.7}
&&\tilde{\hat\Pi}^{(\beta)}=\frac{1}{4\pi}\partial_0\beta^{-1}
+i\lambda B\beta^{-1},\nonumber\\
&&\tilde{\hat\Pi}^{(\beta')}=\frac{1}{4\pi}\partial_0\beta'^{-1}
-i\lambda {\beta'}^{-1}B',\nonumber\\
&&\tilde{\hat\Pi}^{(\tilde V)}=-\frac{1+c_V}{4\pi}\partial_0
\tilde V^{-1},\nonumber\\
&&\tilde{\hat\Pi}^{(\tilde V')}=-\frac{1+c_V}{4\pi}\partial_0
\tilde V'^{-1}.\ear
Canonical quantization then implies the Poisson algebra (see
ref.\cite{AR2,AAR} for derivation; $g$ stands for a generic
WZW field of level $n$)
\bear\label{3.8}
&&\{g_{ij}(x),\hat\Pi^{(g)}_{kl}(y)\}_P=\delta_{ik}\delta
_{jl}\delta(x^1-y^1), \nonumber\\
&&\{\hat\Pi_{ij}^{(g)}(x),\ \hat\Pi^{(g)}_{kl}(y)\}_P
=-\frac{n}{4\pi}\left(\partial_1g^{-1}_{jk}g^{-1}_{li}-g_{jk}
^{-1}\partial_1g^{-1}_{li}\right)\delta (x^1-y^1). \ear
In terms of canonical variables, we have for the constraints
(\ref{2.36}), (\ref{2.46}),
\bear\label{3.9}
&&\Omega_+=-i\tilde{\hat\Pi}^{(\tilde V)}\tilde V-
\frac{(1+c_V)}{4\pi}\tilde V^{-1}
i\partial_1 \tilde V+j_+,\nonumber\\
&&\Omega_-=i\tilde V'\tilde{\hat\Pi}^{(\tilde V')}+
\frac{(1+c_V)}{4\pi}\tilde V'i\partial_1\tilde V'^{-1}+j_-\ear
and for the constraints (\ref{3.5}), (\ref{3.6}),
\bear\label{3.10}
\hat\Omega_-&=&i\beta\tilde{\hat\Pi}^{(\beta)}+i\tilde V\tilde
{\hat\Pi}^{(\tilde V)}-\frac{1}{4\pi}\beta i\partial_1\beta^{-1}
\nonumber\\
&&+\frac{1+c_V}{4\pi}\tilde V i\partial_1\tilde V^{-1}
+\{\hat b_-^{(0)},\hat c_-^{(0)}\},
\ear
\bear\label{3.11}
\hat\Omega_+&=&-i\tilde{\hat\Pi}^{(\beta')}\beta'-i\tilde
{\hat\Pi}^{(\tilde{V'})}\tilde V'\nonumber\\
&&+\frac{1}{4\pi}{\beta'}^{-1} i\partial_1\beta'
-\frac{(1+c_V)}{4\pi}\tilde {V'}^{-1} i\partial_1\tilde V'
+\{\hat b_+^{(0)},\hat c_+^{(0)}\} .
\ear
With the aid of the Poisson brackets (\ref{3.8}) it is straightforward
to verify that $\hat\Omega^a_+=tr(\hat\Omega t^a)$ and
$\hat{\Omega}^a_-=tr(\hat\Omega t^a)$ are
first class:
\be\label{3.12}
\left\{\hat\Omega_\pm^a(x),\hat\Omega^b_\pm(y)\right\}_P=-f_{abc}
\hat\Omega^c_\pm\delta(x^1-y^1). \ee
Hence the corresponding BRST charges are nilpotent. Similar
properties are readily established for the remaining operators
$\Omega_\pm$. Furthermore,
\bear\label{3.13}
&&\{\Omega_+(x), \hat\Omega_-(y)\}_P=0,\nonumber\\
&&\{\Omega_-(x), \hat\Omega_+(y)\}_P=0.
\ear

The physical Hilbert space of the non-local formulation of
$QCD_2$ is now obtained by solving the cohomology problem
associated with the BRST charges $Q_\pm,\hat Q_\pm$ in the
ghost-number zero sector. The solution of this problem is
suggested by identifying this space with the space of
gauge-invariant observables of the original theory defined
by (\ref{2.1}). It is interesting to note that the
constraints $\hat\Omega_\pm\approx 0$ are implemented
by any functional of $V$ (and the fermions), thus implying
that $\tilde V,\beta(\tilde V', \beta')$ can only occur
in the combinations $\beta^{-1}\tilde V(\tilde V'{\beta'}^{-1})$.
Indeed, making use of the Poisson brackets (\ref{3.8}), we have
\bear\label{3.14}
&&\left\{\hat\Omega_-^a(x),\ \beta^{-1}(y)\right\}_P=
i(\beta^{-1}(x)t^a)\delta(x^1-y^1), \nonumber\\
&&\left\{\hat\Omega_-^a(x),\ \tilde V(y)\right\}_P=
-i(t^a\tilde V(y))\delta(x^1-y^1), \nonumber\\
&&\left\{\hat\Omega_+^a(x),\ \tilde V'(y)\right\}_P=
+i(\tilde V'(x)t^a)\delta(x^1-y^1), \nonumber\\
&&\left\{\hat\Omega_+^a(x),\ {\beta'}^{-1}(y)\right\}_P=
-i(t^a{\beta'}^{-1}(y))\delta(x^1-y^1). \ear

As for the other two constraints, $\Omega_+\approx 0$
and $\Omega_-\approx0$ linking the bosonic to the free
fermion sector, they tell us in particular, that local fermionic
bilinears should be constructed in terms of free fermions and the
bosonic fields as follows:
\be\label{2.46a}
\left(\psi_1^{(0)\dagger}\beta^{-1}\tilde V\psi_2^{(0)}\right)
=\left(\psi_1^{(0)\dagger}\tilde V'{\beta'}^{-1}\psi_2^{(0)}\right)
=\left(\psi_1^{(0)\dagger}V\psi_2^{(0)}\right)
=\left(\psi_1^\dagger\psi_2\right).\ee
This is in agreement with our
expectations.

\section{The $QCD_2$ Vacuum revisited}
\setcounter{equation}{0}

The constraints, $\Omega_+\approx 0$
and $\Omega_-\approx 0$, link the $\tilde V$-free fermions-ghosts
and $\tilde V'$-free fermions-ghosts sectors respectively. They operate
in the topological sector associated
with the coset $U(N)_1/SU(N)_1$.

Before proceeding to the solution of the cohomology problem in this sector,
one comment is in order concerning the factorization of the $U(1)$ degree of
freedom.
In fact, the factor
$Z_{coset}=Z^{(0)}_F Z^{(0)}_{gh} Z_{\tilde V}$
in (\ref{2.16}), corresponds
to the partition function of the coset
$U(N)/SU(N)_1=U(1)\times SU(N)_1/SU(N)_1$ \cite{NS}.
By bosonizing the free fermions \cite{WZW} one can factorize the
$U(1)$ degree of freedom, which shows that
it merely acts as a spectator. (This factorization can no longer
be done in the case of more than one flavor, leading to higher level
$SU(N)$ affine Lie algebras).

The solution of the cohomology problem for the topological
coset
\break\hfill
 $SU(N)_1/SU(N)_1$ leads to the existence of $N$
inequivalent vacua \cite{AGSYS}. Each of these can be associated
with a $SU(N)_1$ primary field. There are $N$ such  primary fields in
the $SU(N)_1$
conformal quantum field theory, each one corresponding to a so-called
integrable representation. The restriction in the number of
the allowed representations arises from the affine (Kac-Moody)
selection rules \cite{GW}.
The construction of such primaries in the $SU(N)_1=U(N)/U(1)$ fermionic
coset theory has been carried out in ref.\cite{NS}.

By further gauging the $SU(N)_1$ group we can show that these
primaries are mapped into primaries of the coset
$SU(N)_1/SU(N)_1$ of conformal dimension zero.
These primaries, acting on the Fock vacuum, create
the different inequivalent vacua of the topological coset theory. For the
$U(N)/SU(N)_1$ coset the conformal dimension of the primaries is different from
zero and is determined by the extra $U(1)$ factor.
They are given in terms of the properly antisymmetrized
product of $p$ fermionic bilinears, $p=1,...,N$,
 which in terms of the decoupled fields read:
\be\label{a}
\Phi_p(z,\bar z)=
:e^{2pi\phi}:
:\psi_2^{(0)\dagger i_1}...
\psi_2^{(0)\dagger i_p}:
:\psi_1^{(0) j_1}...\psi_1^{(0) j_p}:
:V_{\cal A}^{-1\ i_1j_1...i_pj_p}: ,
\ee
where
\be\label{d}
V_{\cal A}^{i_1j_1...i_rj_r}\equiv
[:v^{i_1j_1}\ldots v^{i_rj_r}:]_{\cal A}
\ee
Here $v$ stands for $\tilde V$ or $\tilde V'$
(depending on the coset in question), and the subscript
${\cal A}$ means antisymmetrization in the left and
right indices, separately.
The conformal dimension of $V_{\cal A}$ is the conformal dimension of an
$SU(N)_1$ primary
field in the representation $\Lambda_p$ whose Young diagram has
$p$ vertical boxes, as given by \cite{KZ}
\be
h_{\Lambda_p}= \bar h_{\Lambda_p}=\frac{c_{\Lambda_p}}{c_V+k}
\ee
where $c_V=N$ for $SU(N)$, $k=1$ and $c_{\Lambda_p}=\frac{p}{2N}
(N+1)(N-p)$, is the Casimir of the representation $\Lambda_p$.
The additional vertex operator  $:e^{2pi\phi}:$  is a result of the
 factorization of the $U(1)$ spectator as explained
above. It should be stressed that this vertex operator
(with conformal dimensions given by $h=\bar h=-p^2/2N$), is crucial
to obtain the correct dimension of the primaries. They are the intertwining
operators linking the N vacua of the conformal sector,
referred to above.

In the non-conformal sector the primaries (\ref{a}) are replaced by the
 properly antisymmetrized
product of $p$ fermionic bilinears,
\be \label{4.1}
\Phi_p(z,\bar z) =
Tr~{\cal{A}} \left(:\psi_2^{\dagger}\psi_1
\psi_2^{\dagger}\psi_1...\psi_2^{\dagger}
\psi_1:
\right) ,
\quad p=1,...,N,
\ee
which in terms of the decoupled fields
are given by (\ref{a}) with the replacement $v \to V$ in (\ref{d}).
The primaries (\ref{4.1}) implement the constraints $\Omega_{\pm} \approx 0$,
and thus create physical states.

If we assume the $QCD_2$ vacuum to lie in the conformal ($\beta = 1$) sector,
then we must conclude that there exists an N-fold degeneracy of the $QCD_2$
ground state. This generalizes
the conclusion of ref.\cite{AR1}, where
this degeneracy has been discussed in some detail for the case of N = 2.

\section{Conclusion}

The main objective of this paper was to clarify the role of the various BRST
symmetries and associated nilpotent charges present in the decoupled formulation
of $QCD_2$.  Our analysis has shown that of the three nilpotent charges
obtained in ref.\cite{CRS} in the non-local formulation, only two are required
to vanish on ${\cal H}_{phys}$. They correspond to $\tilde \Omega_+ \approx 0$
 and
$\Omega \approx 0$ as given by equations (3.13) and (3.18) of that reference, or
$\Omega_+ \approx 0$ and $\hat \Omega_- \approx 0$  in the notation of this
paper.
These constraints are
first class. We have further shown that the constraint $\tilde\Omega_-$ in
 eq.(3.13) of \cite{CRS} is to be replaced by the constraints
$\Omega_- \approx 0, \hat\Omega_+ \approx 0$ in the present notation. These
again
 represent a first class system. All these constraints were found to be
implemented consistently by suitable products of fermion bilinears,
corresponding to gauge invariant observables of the original partition function
(\ref{2.1}). This solves the corresponding cohomology problem in the
 ghost number zero sector.

The constraints $\hat\Omega_+ \approx 0$ and $\hat\Omega_- \approx 0$
couple the conformal sector of the theory to the sector of massive excitations
$\beta$ and $\beta'$, whose dynamics is described by the partition function
$Z_\beta$ and $Z_{\beta'}$ in (\ref{2.18}) and (\ref{2.42}), respectively.
Assuming that the $QCD_2$ ground state lies in the zero-mass, conformal sector,
one is led to the conclusion that it is
 N-fold degenerate in the case of
an SU(N) gauge symmetry with one flavor. This is in accordance with the
conclusion reached in ref.\cite{AR1}, but is valid only,
provided $\beta$ and $\beta'$
act as identity operators in this sector.

\vspace{1cm}

{\it Acknowledgements:} D.C.C. would like to thank the Deutscher
Akademischer Austauschdienst for the financial
support which made this collaboration possible.

\end{document}